\documentclass[onecolumn,aps,floatfix,notitlepage,superscriptaddress,10pt]{revtex4-1}
\usepackage{amssymb,amsmath,amsfonts}
\usepackage{graphics,dcolumn,bm,epic,eepic,float} 
\usepackage{epstopdf} 
\usepackage{bm} 
\usepackage{color} 
\usepackage{framed}
\usepackage{subcaption}
\usepackage{graphicx}
\usepackage{boxhandler}
\usepackage{etoolbox}
\usepackage{ragged2e}

\begin{document}

\title{Collective Prediction of Individual Mobility Traces with Exponential Weights}

\author{Bartosz Hawelka}
\affiliation{Department of Geoinformatics - Z\_GIS, University of Salzburg, Austria}

\author{Izabela Sitko}
\affiliation{Department of Geoinformatics - Z\_GIS, University of Salzburg, Austria}

\author{Pavlos Kazakopoulos}
\affiliation{CS Research Foundation, Amsterdam, The Netherlands}

\author{Euro Beinat}
\affiliation{CS Research Foundation, Amsterdam, The Netherlands}
\affiliation{Department of Geoinformatics - Z\_GIS, University of Salzburg, Austria}

\date{\today}

\maketitle




{\bf We present and test a sequential learning algorithm for the short-term prediction of human mobility. This novel approach pairs the Exponential Weights forecaster with a very large ensemble of experts. The experts are individual sequence prediction algorithms constructed from the mobility traces of 10 million roaming mobile phone users in a European country. Average prediction accuracy is significantly higher than that of individual sequence prediction algorithms, namely constant order Markov models derived from the user's own data, that have been shown to achieve high accuracy in previous studies of human mobility prediction. The algorithm uses only time stamped location data, and accuracy depends on the completeness of the expert ensemble, which should contain redundant records of typical mobility patterns. The proposed algorithm is applicable to the prediction of any sufficiently large dataset of sequences.}
\hfill \break

The problem of algorithmic prediction of human mobility has received significant attention in the literature in recent years, for its potential applications and its inherent theoretical value. The problem is posed as asking for a prediction of the short-term future location of an individual, given his or hers previous locations and possibly other side information. One can distinguish the algorithms that have been studied in the mobility prediction literature in two classes. On one hand, we have algorithms that use only the single user's past locations, without any other information, to estimate the next location. This individual sequence prediction is closely related to lossless compression  of sequential data \cite{barab_limits, rissanen, feder1992universal}. The other category comprises methods that take advantage of data beyond the user's own past locations (e.g. the network of connections between devices as a proxy of the user's social interactions \cite{DeDomenico2013798}). In this paper, we study an algorithm that falls into the latter category, but uses elements from the former. It combines individual sequence prediction algorithms constructed from the mobility patterns of all users in order to predict the next location of single users. The collective exploration of space from all users in the dataset is tapped to provide predictions for individual users. The proposed algorithm falls in the framework of sequential learning prediction with experts (\cite{plgames_book} and references therein), a family of machine learning algorithms that is well developed theoretically and has been applied to real-life prediction problems in various fields e.g. \cite{cutland1991universal, cohensinger, borodin2000competitive, monteleoni2003online, stoltz2005internal, experts_electricity, mallet2009ozone, vovk2009prediction, dashevskiy2011time, jacobs2011adapting, mallet2010ensemble, monteleoni2011tracking, onlineportfolio}. 

In general terms, the algorithm predicts the next status of an agent in a system where the sequence of previous states of that agent are represented as time-stamped symbols from a finite set. Instead of deriving the next symbol from the agent's sole history, the algorithm looks at the future states predicted by all other agents in the system and gives more weight to those that, in the past, best predicted the future of the agent in focus. Learning and weight adaptation take place at every prediction step to minimise the prediction error. 

Given time-stamped location data of users in an area, one can construct from the mobility trace of each user an individual sequence prediction algorithm, based only on that user's past locations. With enough user traces available, one can make use of these algorithms, which represent, for a sufficiently large dataset, a redundant collection of typical mobility behaviours in the area, to predict individual mobility sequences. When predicting the next location in a users's mobility sequence, individual sequence prediction algorithms derived from \emph{other users} are queried to provide a prediction for the user in focus. From these predictions of individual algorithms, a choice is made by an Exponential Weights (EW) forecaster algorithm at each round to produce a single prediction. The choice of the forecaster is random, with probability proportional to the weight of each expert. After the real next location is revealed, the forecaster promotes experts that gave a correct prediction, and demotes those that are wrong. This is achieved by multiplying the weights of erroneous experts with a learning factor $\beta=e^{-\eta}$, where $\eta>0$ is called the \emph{learning rate}, while leaving the weights of correct experts unchanged for the next round. The algorithms queried by the forecaster are called \emph{experts}, and their collection an \emph{expert ensemble}. To extract the experts from the mobility sequences, we use an $O(1)$ Markov model, constructed from the expert's location data so that the transition probabilities are those seen in the expert's mobility sequence. Hence only a fraction of experts, those that contain relevant transitions, will be able to provide a prediction at any given time step. For mobility at the scale of a country this fraction is in fact typically rather small. Here we study a variant of the EW forecaster, an adaptation for so-called sleeping experts, i.e. for situations where not all experts can make a prediction available to the forecaster at every step  \cite{blum1997empirical, fss, bm, experts_electricity, kleinberg2010regret} (\emph{specialised} experts is an equivalent term used in the literature). For brevity we refer to this variant simply as the EW forecaster, or just EW. 

In the present study, we use the algorithm with anonymized mobile phone call detail records (CDR) as the basic data unit. The mobile phone device is viewed as a proxy of the user's location, and spatial accuracy is defined by the coverage area of the principal antenna cell connected to the device. The data is transformed into time-stamped sequences of characters, each character corresponding to an antenna on the ground. The term \emph{sequence} in this context refers to the mobility trace of a user. In our case the data consists of a sample of anonymized CDRs from more than 10 million roamers in a European country over a period of seven months, made available for the study by a major telecom operator. CDRs have been used in studies of human behaviour (see e.g. \cite{sandy}), and often as proxies of an individual's location in studies of human mobility prediction \cite{barab_limits, haiti2012, ivory2013}. The interest in using CDRs to study human mobility has recently received a new urge in an effort to better understand and slow down the spread of disease outbreaks \cite{Tizzoni2014, eubank2004modelling, tatem2009use, wesolowski2012quantifying}.

In order for the method to provide accurate predictions using the expert/forecaster method, it is a necessary condition that the dataset contains redundancies, i.e. that the users' traces cover the area under study sufficiently. Otherwise, there will often be cases where none of the experts are able of contributing a prediction, and accuracy will be hit. When this basic condition is satisfied however, our approach is particularly suited for predicting mobility traces that can be non-stationary, without the need for additional data other than the time-stamped locations of users. As we will see in our test set, consisting of the thousand longest continuous mobility sequences in the seven month dataset, the EW forecaster outperforms the standard for individual sequence human mobility prediction, the Markov model \cite{song_wifi, ivory2013}. This can be advantageous in applications where the mobility of transient populations is studied, for example populations rapidly changing their mobility patterns as a result of a natural disaster \cite{haiti2012} or a large event, or the mobility of tourists.

\section*{Results}

\subsection*{Telecom dataset} 
The data was provided by a major telecom operator and consists of an anonymised sample of seven months of roamers' CDRs in a European country. The data covers the period between beginning of May to end of November 2013. Each CDR contains the principal antenna that a mobile device is connected to during a phone call, SMS communication or data connection. The time-stamped connection event is interpreted as a location measurement, positioning the device inside the approximate coverage area of the principal antenna. The size of this coverage area can range from a few tens of meters in a city to a few kilometers in remote areas. We do not at all consider the actual geographical positions of the antennas, instead we take this correspondence as a given and represent each antenna as a Unicode (utf-8) character (the total number of antennas in our sample is over $30\;000$, hence the use of the Unicode set for the character assignment). The series of connections of a roaming user is transformed into a time-stamped character sequence $(X_1, X_2,\ldots,X_N)$, which is the object passed to the prediction algorithm.

We take the time step of the sequence of locations to be 1 hour. If more than one events fall within a single time step, one of them is chosen in random to represent the location of the user. In this manner, the mobility trace of a user is converted to an abstract sequence of symbols that unfolds in one-hour steps, and predictions are given for the next hour location, i.e. the next symbol in the sequence. The length of the time step is chosen to balance precision with completeness of the sequences. The sequences of antenna connection events for most users are discontinuous and sparse, having the usual erratic profile of mobile activity patterns. A shorter time unit increases trace fragmentation, while a longer unit would reduce the accuracy of the representation of the actual mobility trace, and consequently the value of the prediction.

The test set of sequences used for benchmarking the algorithm is selected by length, so that we can gather enough statistical data on the behaviour of the algorithm. Specifically, the 1000 longest continuous character strings, i.e. mobility trace fragments, were selected. To make the comparison with individual sequence prediction sharper, we remove from the dataset the rest of the data of the users that have one or more sequences in the set of 1000. In this way it is as if the users in the test set is observed for the first time. The remaining data is used to construct the expert ensemble. Each user's trace is turned into a $O(1)$ Markov model which then plays the role of an expert providing predictions to the forecaster whenever possible. Our sample comprises of more than 10 million users, each of which contributes a prediction algorithm to the expert ensemble. This is an unusually large number of experts. We could find a comparable expert ensemble used only in \cite{cohensinger}, in the context of document classification. This unusual abundance of experts is not incidental, instead it defines our approach. We exploit the redundancy of mobility patterns in our dataset to predict traces effectively.

We present the results of the empirical tests of the forecasters from two complementary aspects. We focus first on measuring the absolute performance of the EW forecaster on our 1000 test sequences. We then present aspects of the internal dynamics of the forecaster, and examine how the performance is affected by varying parameters that modify the content of the expert ensemble.

\subsection*{Prediction accuracy}
Human mobility is characterised by strong regularities that make it possible in principle to predict it with high precision \cite{barab_limits, haiti2012, ivory2013, gonzalez2008}. It does not however always exhibit these regularities. There are contexts in which human mobility patterns either change abruptly or are by their nature non-repetitive. Our test data of mobile network roamers exemplifies this. Tourists and foreign visitors in general are naturally expected to be less regular in their patterns than local residents, in some cases merely crossing through a country without any repetitive patterns at all. In a different sort of situation, when facing an emergency, mobility behaviours can change in a short time, usually while transitioning between more regular regimes \cite{haiti2012}. Transient mobility behaviour is also expected at the individual level as a result of changes of daily routines. Sequential learning algorithms are exactly designed to address problems where the nature of the predicted sequence cannot be accounted for a priori, or even assumed to be persistent in time. A diversified approach using many experts overseen by a forecaster is then a better strategy for accurately predicting diverse types of sequences, in our case mobility traces.

The predictive power of the forecaster/expert combination we study is drawn from the redundancies that characterise large mobility datasets. When a single user following a new, perhaps non-repetitive pattern, e.g. while vacationing, their past position data cannot inform short-term predictions. From the point of view of a dataset containing the traces a sufficiently large and diverse set of users however, patterns that are new or irregular for the individual can be found to have already been traced by other users. In the case of tourists in particular, itineraries that are a first for a user have most often already been explored by others in part or in whole. These past patterns are encoded in individual prediction algorithms and, combined by a forecaster, can provide accurate predictions during a transient phase without depending on the regularities of a single individual. In addition, the expert ensemble can be dynamic, with experts added or removed on the fly, e.g. choosing the experts in a moving time window.

\hfill \break
\begin{figure}[h!]
\captionsetup{justification=centerlast}
\centering
\begin{subfigure}{.5\textwidth}
  \centering
  \includegraphics[width=0.8\linewidth]{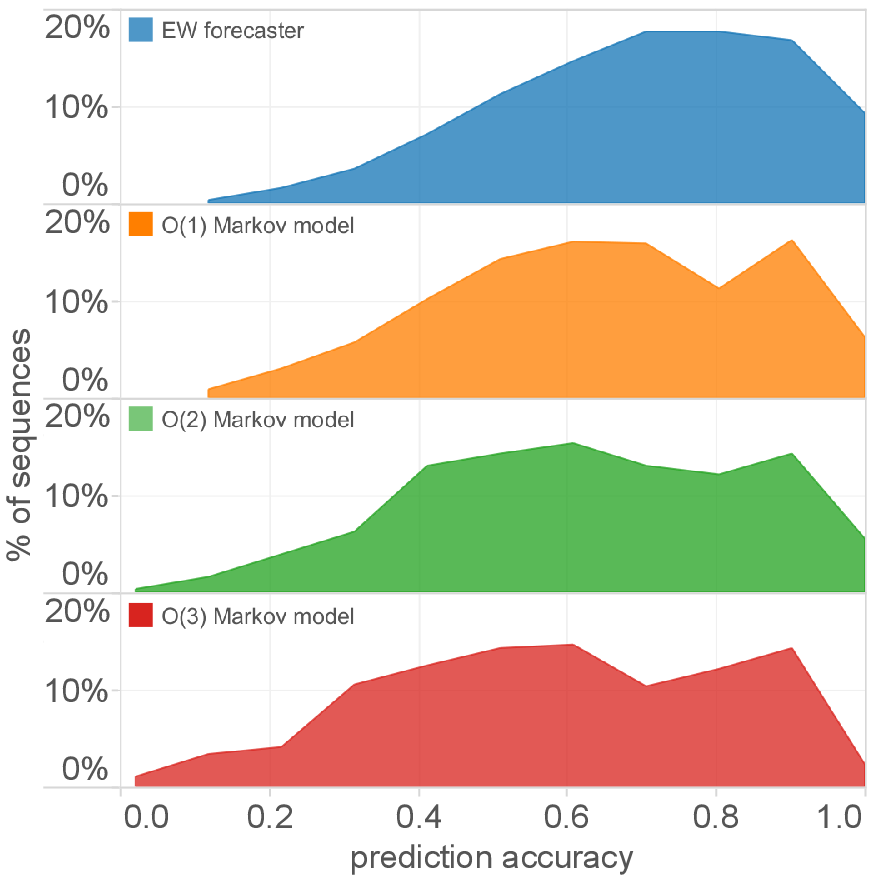}
  \caption{}
  \label{us_Markov123_bins}
\end{subfigure}%
\begin{subfigure}{.5\textwidth}
  \centering
  \includegraphics[width=0.8\linewidth]{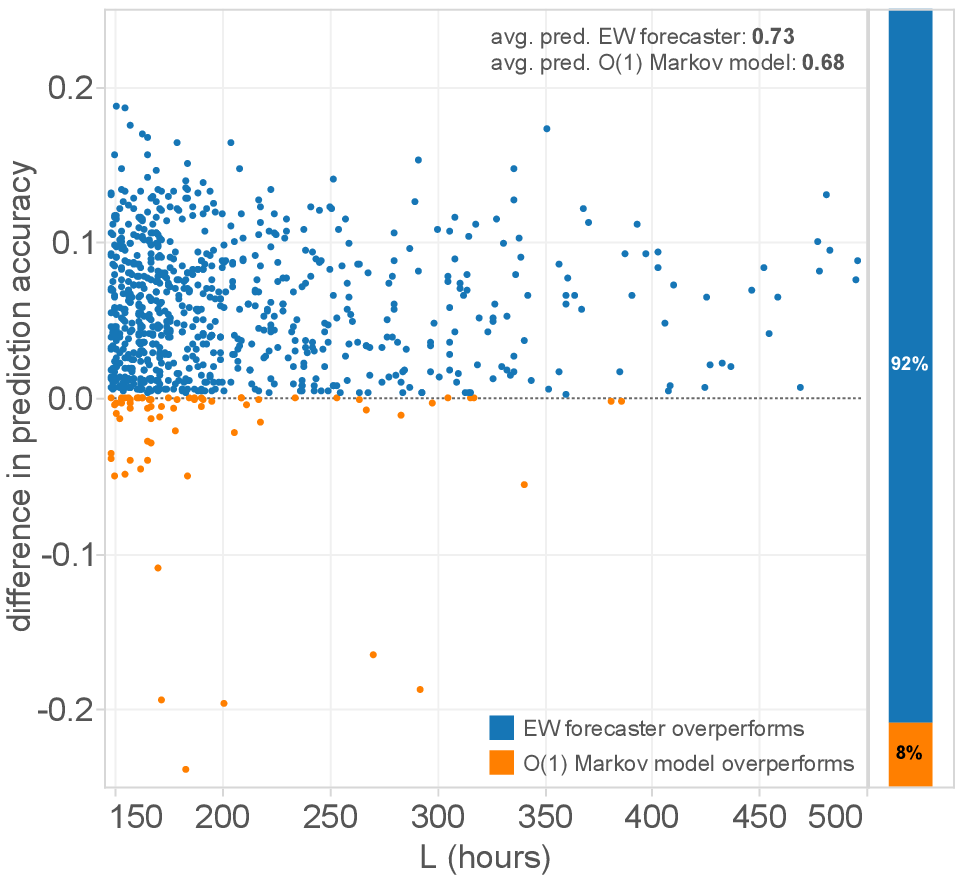}
  \caption{}
  \label{diff_us_Markov} 
\end{subfigure}
\caption{\textbf{EW forecaster prediction accuracy.} (a) Percentage of sequences predicted with a certain accuracy (in bins of $10\%$) for the EW forecaster and Markov models of order $k=1,2,3$ constructed sequentially from the users own data as the sequence of locations is observed in time. We use a learning rate $\eta = 3$. The EW forecaster improves on the performance of the best Markov model, which again turns out to be $O(1)$ \cite{song_wifi, ivory2013}, by an average of $5\%$. A detailed comparison between the two is depicted in (b), the scatterplot of difference in prediction accuracy per sequence. For more than $90\%$ of the test sequences, the EW forecaster is more accurate.}
\label{1ab}
\end{figure}

We measure the performance of the EW sleeping experts forecaster on our set of test sequences. We compare its performance with an important benchmark for human mobility prediction at this resolution and duration, the $O(k)$ Markov model. The expert ensemble is defined (per sequence) by admitting only expert sequences that ended in a fixed time $T_{past}$ before the predicted sequence starts. For the comparative testing we took $T_{past}$ to be 3 months ($2160$ hours). We also admit all experts in our sample, without any filters. This setup provides optimal performance.

\begin{figure}[h!]
\captionsetup{justification=centerlast}
\centering
\begin{subfigure}{.5\textwidth}
  \centering
  \includegraphics[width=0.8\linewidth]{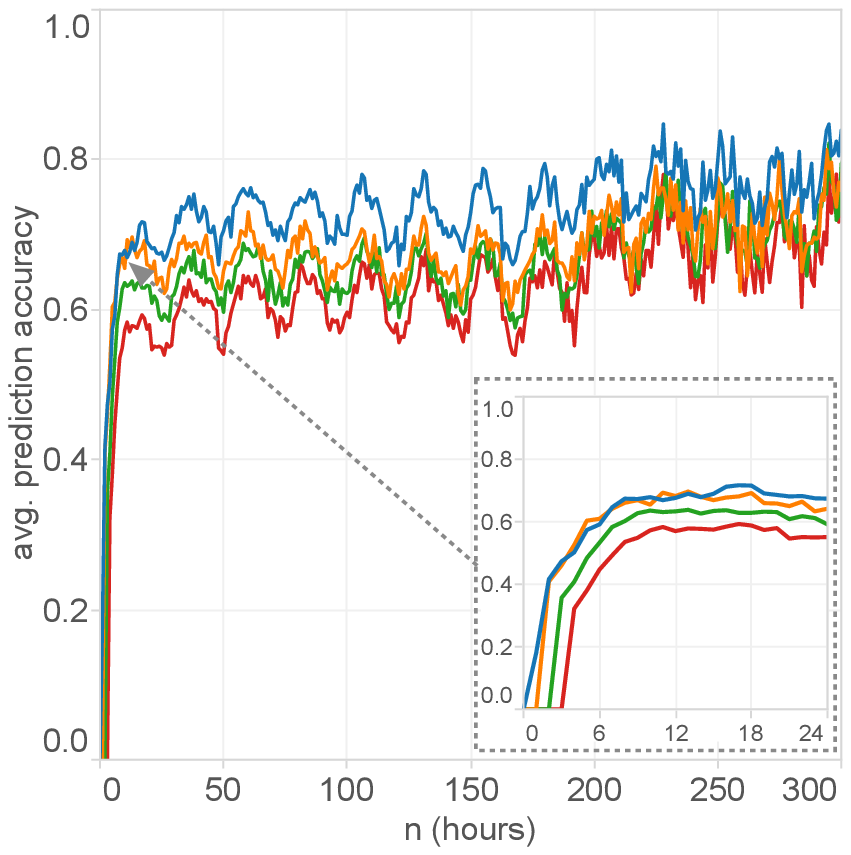}
  \caption{}
  \label{us_Markov123_per_pos}
\end{subfigure}%
\begin{subfigure}{.5\textwidth}
  \centering
  \includegraphics[width=0.8\linewidth]{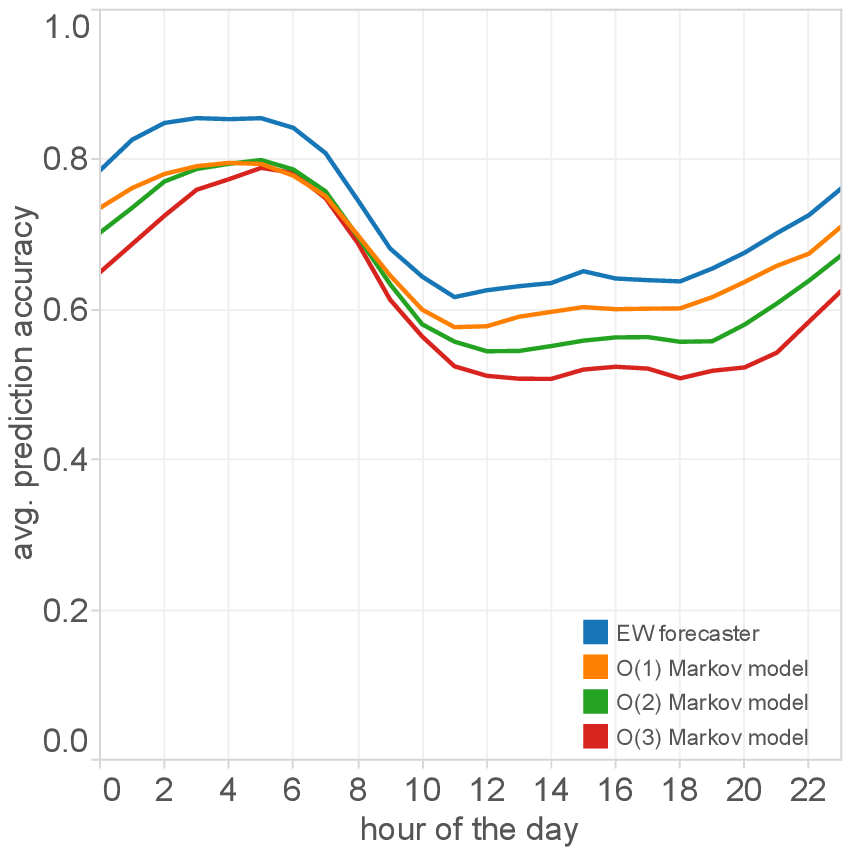}
  \caption{}
  \label{pred_over_day}
\end{subfigure}
\caption{\textbf{Prediction per position and over a hour of a day.} (a) Average prediction accuracy per position $n$ in the sequence, for the EW forecaster and Markov models  orders $k=1,2,3$. The best Markov model is $O(1)$  and is on par with the EW forecaster for the first half-day after the start of the user's sequence and the prediction process. EW achieves a stable (average) lead after that point. The quasi-periodic pattern is due to the fact that most roamers arrive to the visit country during the day, combined with the fluctuation between day and night prediction accuracies seen in (b). Prediction accuracy is significantly higher in the period between 02:00 - 08:00 because of the much higher regularity of mobility patterns during these hours.}
\label{2ab}
\end{figure}

The performance of the EW forecaster can be seen in Fig.~\ref{1ab}, in comparison with Markov model individual sequence predictors of order $k=1, 2, 3$. The best settings for the EW forecaster turn out to be at a value $\eta\approx 3$, but the accuracy is only slightly higher than in the adaptive version (Supplementary Information S2), where the learning rate $\eta$ is free to vary over a grid of values, and at each step is given the median of the values of $\eta$ that have had the best performance so far. In the diagrams we always use $\eta = 3$. In addition, the exclusion of the user's own location sequence (as it unfolds in time) in the expert ensemble gives a slightly better average prediction accuracy than one gets when admitting it as an additional expert. As we can see, the EW forecaster performs significantly better than Markov models, with the $O(1)$ model being the most accurate among the latter. This is consistent with the results seen in previous studies of human mobility prediction. Our choice of order 1 for the Markov models of the experts is based on this ranking. The distribution of the differences in accuracy for individual test sequences shows that EW gives a 5\% average advantage over the $O(1)$ Markov model. When predicting a new sequence, EW overtakes the Markov models in accuracy after an average of 14 hours [Fig.~\ref{us_Markov123_per_pos}]. The quasi-periodic pattern in [Fig.~\ref{us_Markov123_per_pos}] is explained by the day-night variation in prediction accuracy [Fig.~\ref{pred_over_day}], combined with the fact that most foreign visitors arrive during the day.

\hfill \break

\subsection*{Internal dynamics of the forecaster}
To understand better when and why the sequential learning algorithm works or not, we examine the internal dynamics of our forecaster/expert ensemble combinations. The sequential learning algorithm considered as a dynamical system evolving in time contains millions of degrees of freedom, namely the experts' weights, and is statistical in nature. Its dynamics are described by an equal number of difference equations that depend on the sequence under prediction, and do not avail any simple treatment. Nevertheless, empirical metrics can shed some light on the factors crucial for prediction accuracy.

\hfill \break
\begin{figure}[h!]
\captionsetup{justification=centerlast}
\centering
\begin{subfigure}{.5\textwidth}
  \centering
  \includegraphics[width=0.8\linewidth]{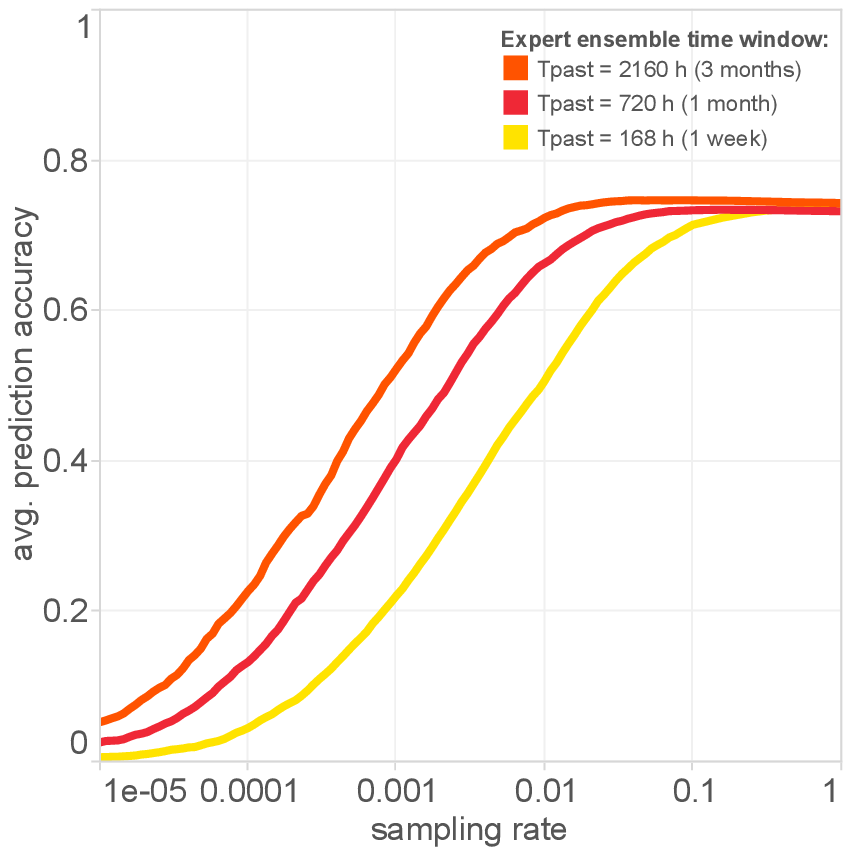}
  \caption{}
  \label{pred_tpast_vs_sample}
\end{subfigure}%
\begin{subfigure}{.5\textwidth}
  \centering
  \includegraphics[width=0.8\linewidth]{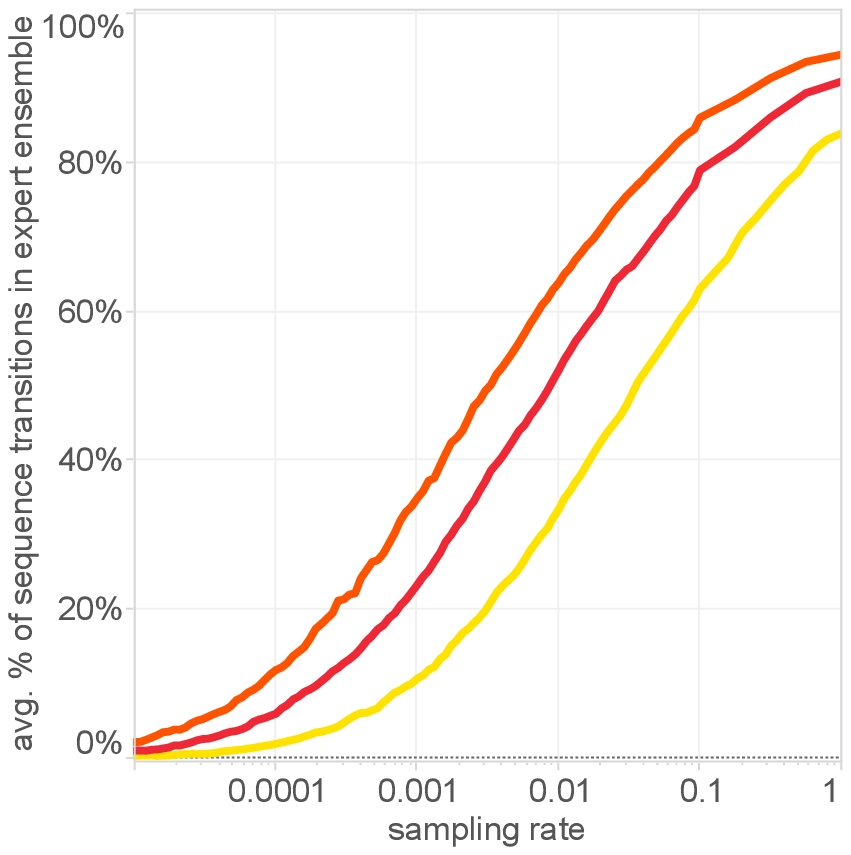}
  \caption{}
  \label{seq_trans_vs_sample}
\end{subfigure}
\caption{\textbf{Prediction accuracy dependence on sampling and $T_{past}$.} (a) Average prediction accuracy for particular filterings of the expert ensemble. We randomly sample experts from the ensemble and additionally we filter the experts' sequence fragments so that only those that end within a time window $T_{past}$ are included. Decreasing the sampling rate and/or reducing $T_{past}$ decimates the ensemble, and beyond a point it hits the accuracy of the forecaster. (b) The average percentage of distinct transitions $X_{n-1}\rightarrow X_{n}$ in a test sequence that are contained by at least one expert in the ensemble after filtering. Prediction accuracy in (a) starts dropping when the sampling rate is reduced beyond a few percent, showing that the ensemble is very diverse and robust. A very slight drop in performance comes with including all experts, due to the logarithmic search costs of the forecaster when the ensemble grows.}
\label{sampling}
\end{figure}

In Fig.~\ref{pred_tpast_vs_sample}, we see the dependence of prediction accuracy from the EW forecaster on two parameters modifying the expert ensemble, a sampling rate at which the ensemble is randomly sampled, and $T_{past}$, the time span before the start of the predicted sequence from which prediction data is admitted to the ensemble. In both cases, performance is stable over a wide range of the parameters and quickly degrades outside. This demonstrates on one hand the robustness of the performance over changes in the expert ensemble and on the other the quick failure when the expert ensemble starts becoming incomplete.  As we see, below a certain threshold of a few percent sampling rate the algorithms performance drops significantly. The fast drop in performance is due to the decimation of the transitions between locations available in the $O(1)$ Markov models of the expert ensemble when experts are filtered out. In Fig.~\ref{seq_trans_vs_sample} we plot the average percentage of unique antenna-to-antenna transitions in the test sequences which are also contained in the expert ensemble, as a function of the sampling rate. The performance of the forecaster depends most crucially on the quality of the ensemble. When the ensemble is diverse, shifting mobility patterns are quickly picked up and correctly predicted by the relevant experts. The EW forecaster significantly promotes their weights relative to other experts over just a few time steps. While the Markov model (or any individual sequence prediction algorithm) has to gather enough statistical information about the new behaviour before producing correct predictions, the forecaster has most of this information already available in the experts.

\begin{figure}[h!]
\captionsetup{justification=centerlast}
\centering
\includegraphics[width=1\linewidth]{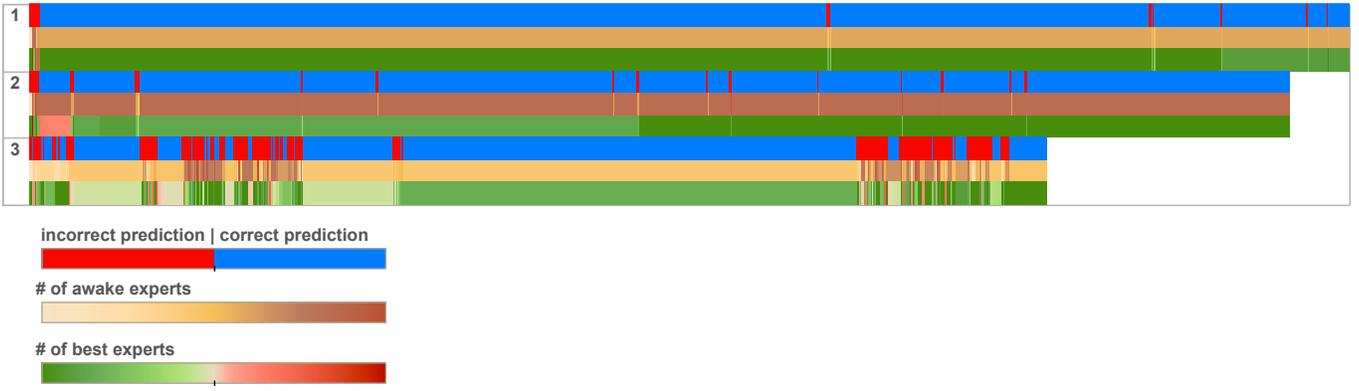}
\caption {\textbf{Correct/incorrect prediction for given position for three selected sequences.} Prediction accuracy in our setup depends crucially on the availability of good experts in the ensemble. In the three example sequences we see a colour-coded depiction of prediction success or failure adjacent to the numbers of awake and best experts, i.e. experts that can provide a prediction at a given step, and those among them which have accumulated the minimum loss up to that step. The three sequences are rather typical examples seen in the test dataset. Low numbers of best and awake experts almost invariably lead to incorrect predictions, and vice versa.} 
\label{example_of_three_seq} 
\end{figure}

This overall dependence on the availability of transitions in the expert ensemble can be also seen when zooming in to single sequences. The three sequences shown in Fig.~\ref{example_of_three_seq} are coloured in three different scales, showing the qualitative correlations between the numbers of best and awake experts and success or failure in prediction. They have been picked to represent three types of prediction dynamics typically seen in our test set. It is clear that the probability of success correlates strongly with the number of best experts at any given time step. This number can stay relatively stable over a segment and then change up or down abruptly as the user moves to a new antenna for which few experts are available. When many experts are available for the current mobility patch of the user, the forecaster quickly starts producing correct predictions.

\begin{figure}[h!]
\captionsetup{justification=centerlast}
\centering
\begin{subfigure}{.5\textwidth}
  \centering
  \includegraphics[width=0.8\linewidth]{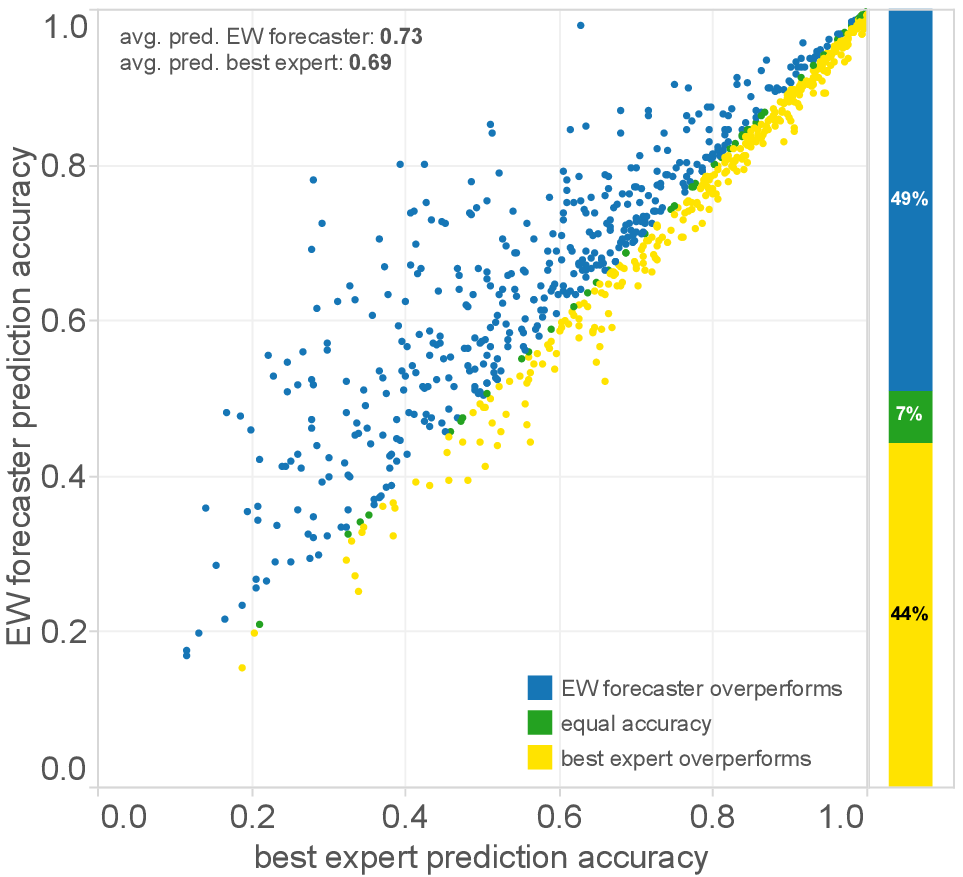}
  \caption{}
  \label{us_vs_best_expert}
\end{subfigure}%
\begin{subfigure}{.5\textwidth}
  \centering
  \includegraphics[width=0.8\linewidth]{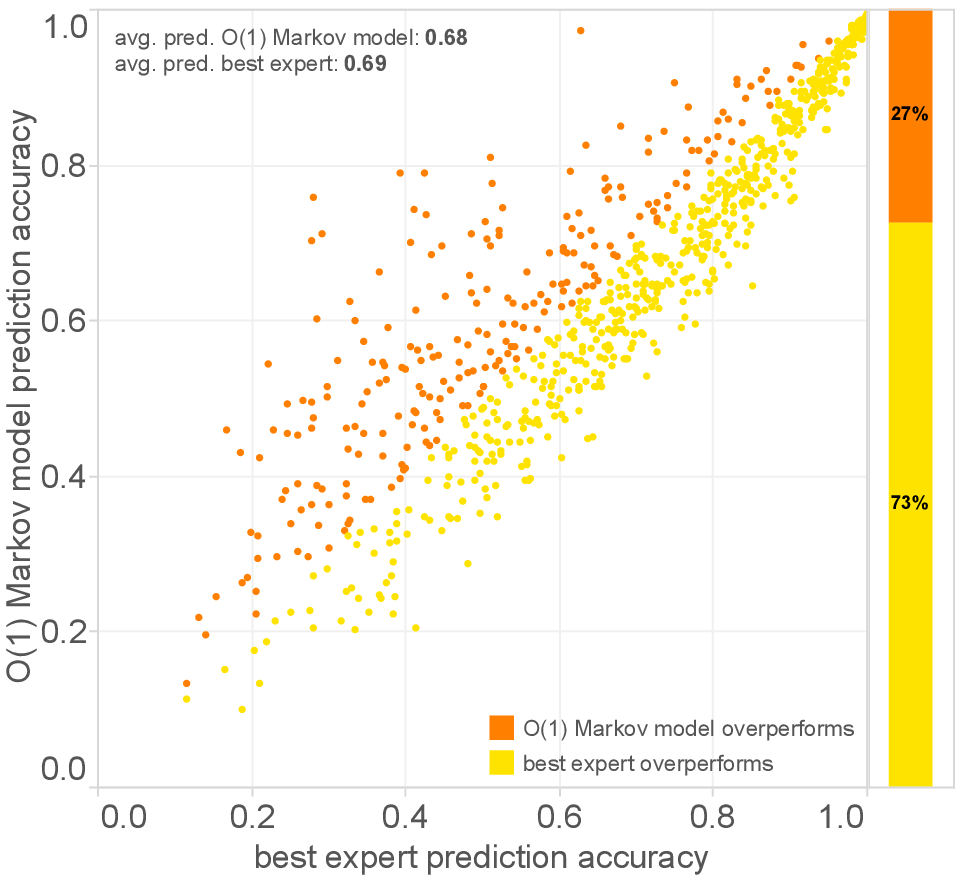}
  \caption{}
  \label{markov_vs_best_expert}
\end{subfigure}
\caption{\textbf{Comparison with the best expert in the ensemble.} The best expert here is declared at the end of the sequence, as the Markov model in the expert ensemble which accumulated the minimum loss during prediction. If more than one experts share this property, a representative is chosen arbitrarily. (a) The EW forecaster's prediction accuracy compared to the best expert prediction accuracy. The forecaster's accuracy is superior more often than not, and with larger differences, resulting in a $4\%$ average advantage. (b) The $O(1)$ Markov model constructed sequentially from the user's own locations as they are recorded in real time is less accurate than the best expert for a large majority of the test sequences. It may appear slightly surprising that another users data is better at predicting a given user's location sequence, but the user's own Markov model is constructed sequentially, needing time to learn the patterns, while experts' Markov models enter the "competition" fully constructed.}
\label{4ab}
\end{figure}

The EW forecaster's internal benchmark is \emph{regret}, the difference in cumulative loss - here the number of erroneous predictions - between itself and the best expert, i.e. the expert with the minimum loss, in those rounds where the expert was awake \cite{fss, bm}. Here we use a different measure, comparing the forecaster's predictive accuracy, with the accuracy obtained by each expert when predicting the sequence alone. For the comparison we declare as \emph{best expert} the one that attains the best prediction accuracy over the whole sequence. In our test set, the forecaster is often significantly more accurate than the best expert, while never performing much worse [Fig.~\ref{us_vs_best_expert}]. The $O(1)$ Markov model constructed sequentially from the predicted user's data in contrast is performing more poorly, with the best expert holding a significant advantage in the majority of cases [Fig.~\ref{markov_vs_best_expert}]. In effect, the mobility trace of a different user is often a better predictor for another user's trace, than the latter's own mobility data. Of course this is due partly to the dynamical construction of the user's own Markov model, which is populated with new transitions as they happen, while the experts' Markov models are derived from past data in a time window $T_{past}=3$ months.

\section*{Discussion} 
We have shown that a large number of individual sequence prediction algorithms derived from the mobility traces of mobile phone users can be combined by an Exponential Weights forecaster to provide accurate next-hour location predictions for individual users. Using a dataset of mobility traces, we have demonstrated the potential of the method to predict short trips of transient populations such as tourists, which, in general, are in general non-stationary. The method can easily be implemented for CDRs, even when the data is highly incomplete, with many gaps in time. It outperforms the Markov model standard for individual sequence human mobility prediction while requiring only time stamped locations at the input.

The proposed method is domain-agnostic, and can in principle be applied to the prediction of any dataset of time series that can be encoded into character sequences.  The only essential requirement is that the sequence dataset contains enough observations so that the phase space of the dynamical system under prediction is already covered, many times over if possible. This is in contrast to individual sequence prediction algorithms, where solely the data of a single sequence is needed to make predictions; it enjoys the advantage of fast adaptation to new mobility patterns for a newly observed agent, or in cases where the event sequence is transient. 

\section*{Methods}
\subsection*{Sequential prediction: experts and forecasters} 
The problem of human mobility prediction is naturally formulated as a sequential prediction problem. The sequence of positions unfolds in time, and the data available for predicting the next position lies in the past. Sequential prediction methods have been developed as a means to provide guarantees on the quality of predictions without making any a priori assumptions on the nature of the sequence. In sequential prediction with experts, the unknown character of the unfolding sequence is anticipated by combining a collection of prediction algorithms instead of a single one. The goal is to make the algorithm able to adapt to non-quasi-stationary transient patterns that may be encountered in the sequence. The benchmark for the quality of prediction is the performance of the best expert, or of an optimal combination of experts. Instead of depending only on the sequence, as in individual sequence prediction when a universal algorithm is used, the best possible prediction rate depends both on the sequence and on the ensemble of experts. 

The absolute performance of an expert is measured by a loss function $l^i(Y_{n}^{i}, X_n)$ that quantifies the difference between the $i$th expert's prediction $Y_{n}^{i}$ and the actual outcome $X_n$ (in our case, the user's position in the next hour, indexed by the positive integer $n$). Among experts in a finite set, there are always one or more that will suffer the minimum loss over a given sequence. One of these experts can be chosen as representative of this class, as the best expert. The individual predictions are combined by a forecaster, an algorithm that assigns to each expert a weight $w_i > 0$ and makes a prediction $Y_n$ for the actual prediction based on these weights. The forecaster's goal is to minimise its own loss: 
\begin{eqnarray}
L_N=\sum^{N}_{n=1}{L(Y_n, X_n)} 
\label{forloss}
\end{eqnarray}
compared to the loss of the best expert in the ensemble, for a sequence of length $N$. This relative loss is called regret and is always measured in hindsight:
\begin{eqnarray}
R_N=L_N-\sum^{N}_{n=1}{l^b(Y_{n}^{b}, X_n)} ,  
\label{regret}
\end{eqnarray}
where $b$ is the index of the best expert, i.e. the expert with the minimum cumulative loss at step $n$. Following the standard in the human mobility literature, we use the simple binary loss function 
\begin{eqnarray}
l^i(Y_{n}^{i}, X_n)=\delta(Y_{n}^{i}, X_n), \mbox{  where  } \delta(A,B) =
\left\{
	\begin{array}{ll}
		0  & \mbox{if } A = B \\
		1 & \mbox{if } A\neq B
	\end{array}
\right.
\label{loss}
\end{eqnarray}
The forecaster loss function $L(Y_n, X_n)$ is also taken to be $\delta(Y_{n}, X_n)$. An additional complication arising in mobility prediction with trace-derived experts is that at every round only a fraction of the experts can provide predictions. The Markov model for an expert is of fixed order (see Supplementary Information S1 for a discussion on the choice of algorithms). It is constructed by counting the relative frequency of transitions between antennas in a user's history. Users generally explore a very small subset of the possible transitions, and so only users that have in the past connected to a given antenna or a sequence of antennas can provide predictions for users currently in the same location. As a result, most experts will abstain from prediction at any given round. When experts cannot provide a prediction at every step, they are called sleeping or specialised. The specific versions of the EW forecaster that we use can be found in Supplementary Information S2. For sleeping experts forecasters, theoretical bounds on regret based on varying definitions have been derived in \cite{blum1997empirical, fss, bm, experts_electricity, kleinberg2010regret}. The bounds compare the forecaster to the best expert or convex combination of experts in those instances where the expert was awake. We do not attempt here to prove a bound for the variant we study. 

The two main ingredients of an sequential learning algorithm are the mechanism for updating the weights after each round, and the rule for combining the experts. Some version of a majority vote is usually chosen for the latter. In our case the prediction outcomes are discrete characters corresponding to individual antennas. At each round the forecaster randomly picks a single prediction out of those provided by the experts with probability proportional to the weight of each expert.  The weight update mechanism is a central factor behind the quality of the predictions, together with the quality of the expert ensemble. Exponential Weights forecasters penalise experts that make wrong predictions by multiplying their weights with a factor less than unity, depending on the loss and a learning rate $\eta > 0$. A higher learning rate accelerates the process of weight update:
\begin{eqnarray}
w_{n+1}^{i}=e^{-\eta l^i(Y_{n+1}^{i}, X_{n+1})}w_{n}^{i}. 
\label{weightupdateew}
\end{eqnarray}
Experts with lowered weight will contribute less in the next round. Those that are often correct will see their predictions selected with higher probability. We test the EW forecaster both with fixed learning rate and an adaptive version where the value of $\eta$ is tuned sequentially as in \cite{experts_electricity}. In addition, we tested versions where the user's own position data, encoded in an individual prediction algorithm of the same type as for the other experts, is added to the expert ensemble. These versions guarantee that in the infinite time limit the expert ensemble contains at least one expert that can reach the performance expected from an individual sequence prediction algorithm, but they consistently performed slightly worse that the corresponding variants where the user's own Markov model was not included.

The central idea of our approach is that with enough users in the dataset, the space of mobility patterns will be densely covered. Users will show similarities in the type and frequency of transitions between antennas. When these transitions are encoded in an individual sequence prediction algorithm, e.g. a Markov model, one user's past data can be useful in predicting another user's future mobility. The bootstrapped predictions draw exclusively from the position data of users without making a priori assumptions about the sequence, such as stationarity (Supplementary Information S3), or requiring additional data sources. The forecaster learns and adapts its choices based only on the success of each expert in providing correct predictions.


\section*{Acknowledgments}
This research was funded by the CS Research Foundation, Amsterdam, The Netherlands (www.collectivesensing.org) and by the Austrian Science Fund (FWF) through the Doctoral College GIScience (DK W 1237-N23), Department of Geoinformatics - Z\_GIS, University of Salzburg, Austria. P.K. would like to thank Panayotis Mertikopoulos for useful discussions.


\end{document}


\title{Collective Prediction of Individual Mobility Traces with Exponential Weights \\ - \\ Supplementary Information}

\author{Bartosz Hawelka}
\affiliation{Department of Geoinformatics - Z\_GIS, University of Salzburg, Austria}

\author{Izabela Sitko}
\affiliation{Department of Geoinformatics - Z\_GIS, University of Salzburg, Austria}

\author{Pavlos Kazakopoulos}
\affiliation{CS Research Foundation, Amsterdam, The Netherlands}

\author{Euro Beinat}
\affiliation{CS Research Foundation, Amsterdam, The Netherlands}
\affiliation{Department of Geoinformatics - Z\_GIS, University of Salzburg, Austria}

\date{\today}

\maketitle


\subsection*{S1. Construction of the experts}
In general, there are no structural constraints in choosing the experts in an sequential learning setup. The guiding principle is that there be good experts in the ensemble. Bad experts, those that give many wrong predictions, are demoted and neutralised by the forecaster. The ensemble can be dynamic, with experts added or removed at will. In this study the experts are derived from the dataset of past user traces. A trace, i.e. a sequence of successive locations of the user sampled every hour, is turned into a prediction algorithm by using the relative frequencies of transitions between locations as transition probabilities in a Markov model of constant order $k$. If $(X_1X_2\ldots X_N)$  is one such trace, where $X_n$ represents the location of the user at time-step $i$ (here time steps are one hour), then an $O(k)$ Markov model can be constructed using as the transition conditional probabilities $P(X_n | X_{n-1}X_{n-2}\ldots X_{n-k})$ the corresponding empirical relative frequencies of the transitions $\ldots X_{n-k}\ldots X_{n-2}X_{n-1} \rightarrow X_n$ as found in the sequence $(X_1X_2\ldots X_N)$. Here we use only $k = 1, 2, 3$,  as these are the values that have emerged as the best choices for human mobility prediction \cite{song_wifi, ivory2013}. Higher values of the order, or variable order Markov models need long quasi-stationary sequences to reach their maximum performance, which in many cases can be impractical for human mobility, e.g. for the prediction of short atypical trips. 

\begin{figure}[h!]
\captionsetup{justification=centerlast}
\centering
\includegraphics[width=0.4\linewidth]{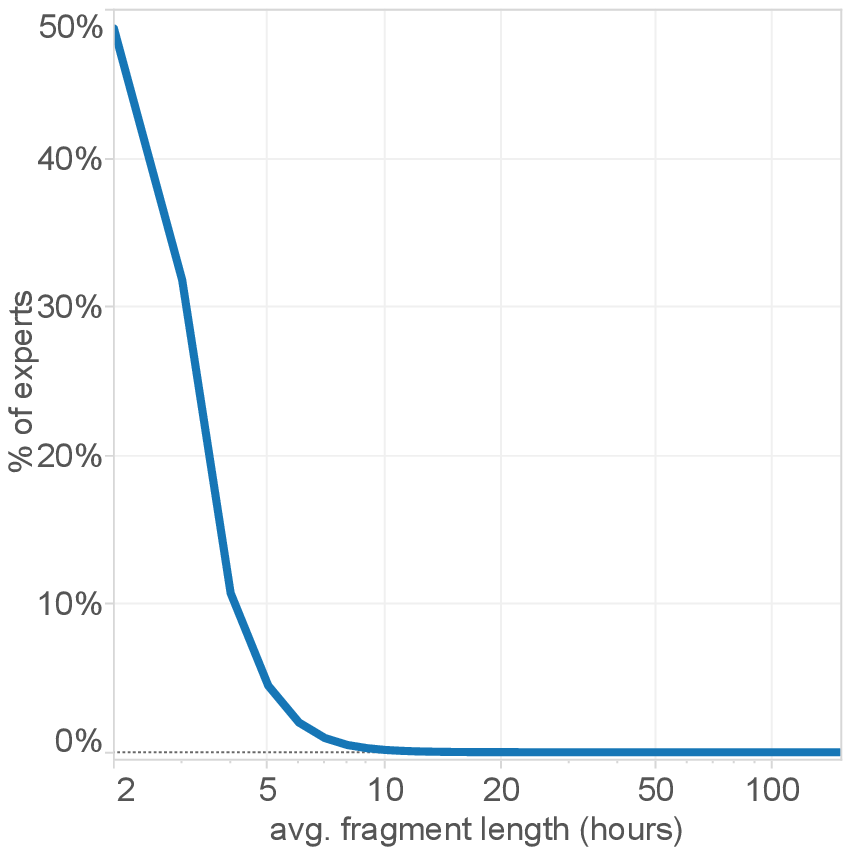}
\caption{\textbf{Experts fragmentation.} Average length of a fragment in an expert sequence. Almost $95\%$ of the experts have an average fragment length of $5$ hours or less. The high degree of sequence fragmentation, due to inherent irregularities in the sequence of connection events in the CDR data, distorts the statistics of the user's Markov model relative to the real frequencies of transitions. Presumably, a more complete record of transitions would increase the prediction accuracy. However, the EW forecaster benefits from the inclusion of all experts in the dataset, even those that are extremely fragmented, indicating that the completeness of the transition record is more crucial to its performance than the accuracy of the Markov models.}
\label{experts_fragments}
\end{figure}

In our dataset, consisting of anonymized call detail records (CDRs) of roamers, the traces of users are in general very fragmented, because of the native irregularities in the activities (calls, SMS, data connection) contained in the CDRs, but also because of multiple visits to the country, or change of preferred roaming network during a visit. We construct one expert per user, simply compiling the statistics for the Markov model from all the available fragments. This of course leads to some distortion of the statistics compared to the real transition frequencies, which can be significant for users with very fragmented frequencies [Supplementary Fig.~\ref{experts_fragments}]. Complete sequences would produce more accurate experts. However, the exponential weight update mechanism ensures that the number of steps required to "discover the good experts" in the ensemble is logarithmic in the number of experts \cite{plgames_book}, and increasing the number of experts by admitting even the most fragmented actually increases the average accuracy marginally. Adding more experts to the ensemble turns out to be beneficial, because it increases the probability that good experts will be found there for a larger class of sequences, while the "search cost" for these experts scales only logarithmically.

\subsection*{S2. Sleeping Experts Exponential Weights forecaster}
Sequential learning algorithms for prediction are based on the idea of using a diversified ensemble of prediction algorithms, or experts, and combining their predictions in every round giving more weight to the predictions of algorithms that have been more accurate so far. To quantify accuracy, a loss function of two arguments, the prediction and the real outcome, is used. Each expert is assigned a weight, which is effectively decreased whenever the the expert provides a wrong prediction. The type of forecaster we use in this study is Exponential Weights, named after the mechanism it employs for updating the expert weights after each prediction round. If $l^i(Y_{n+1}^{i}, X_{n+1})$ denotes the cost function of the $i$th expert, where  $Y_{n+1}^{i}$ is the expert's prediction for the next sequence element at step $n-1$,  and  $X_{n+1}$ the actual next element, then a wrong expert's weight is multiplied with the factor $\beta_{n}=e^{-\eta l^i(Y_{n+1}^{i}, X_{n+1})}$, where $\eta > 0$ is the learning rate at that step. There are many variants of the EW forecaster, their use depending on the constraints of the problem and the objective. In our case, experts are not always able to provide a prediction. The experts are Markov models extracted from mobility traces, which can make a prediction for another trace only when there is some overlap in locations that were visited by the users. Hence we employ the so-called sleeping (or specialised) experts variant \cite{blum1997empirical,fss,bm,experts_electricity,kleinberg2010regret}. At a prediction round $n$, if $E$ denotes the ensemble of experts, only a subset $A_n\subseteq E$ can provide a prediction, and these are the awake experts in this round. The rest of the experts are sleeping, and their weights are not altered. The rule for combining the outcomes predicted by the awake experts into a single prediction at every round is typically some version of the majority rule. In our case it is chosen randomly among the awake experts' predictions, with the probability proportional to the weight of the expert. The procedure is summarised in the box below:

\begin{framed}
\justify
Initialize weights for each expert $i$: $w_{0}^{i}=\frac{1}{|E|}$
\justify
At step $n=1\ldots N$: 
\justify
1) Collect predictions from awake experts $Y_{n+1}^{i}$, $i\in A_n$, according to the $O(k)$ Markov model transition probabilities $P_i(Y_{n+1} | X_{n}X_{n-1}\ldots X_{n-k+1})$.\\
\\
2) Pick randomly a single $Y_{n+1}$ among the $Y_{n+1}^{i}$ with probability $\frac{w_{n}^{i}}{\sum\limits_{i\in A_n}^{} w_{n}^{i}}$. \\
\\
3) Observe the true outcome and calculate the loss function value of each awake expert $l^i(Y_{n+1}^{i}, X_{n+1})=\delta(Y_{n+1}^{i}, X_{n+1})$, $i\in A_n$.\\
\\
4) Update the weights of the awake experts $w_{n+1}^{i}=e^{-\eta l^i(Y_{n+1}^{i}, X_{n+1})}w_{n}^{i}$, $i\in A_n$. Leave the weights of sleeping experts unchanged $w_{n+1}^{i}=w_{n}^{i}$, $i\notin A_n$.\\

\end{framed}

The learning rate $\eta$ is the only free parameter, and can vary adaptively along the sequence, following some additional rule that defines the adaptation. A higher learning rate means that the weights of awake experts that are wrong in their prediction are reduced more. We tested the forecaster with both constant and varying $\eta$. The mechanism of adaptation we used follows the ideas in \cite{experts_electricity}. A logarithmic grid of 30 points equally log-spaced between $10^{-2}$ and $10^{3}$ was chosen for $\eta$, and at each step the forecaster used the median of the grid value that had attained the best accuracy so far. In the non-adaptive version, the same grid value was used for all sequences at every step, and the value providing the best average accuracy is chosen in hindsight. The average prediction accuracy as a function of $\eta$ displays the behaviour shown in Supplementary Fig.~\ref{pred_per_eta}. We see that the forecaster attains its best average performance around $\eta \approx 3$, and higher values give essentially the same result. This is a result of the very large size of the expert ensemble. Experts are typically so abundant at every step that it is a better strategy for the forecaster to effectively eliminate them from subsequent rounds even after a single erroneous prediction.  The distribution of the value of the optimal learning rate $\eta$ that gives the best prediction accuracy (Supplementary Fig.~\ref{best_eta_dist}) shows that almost all sequences are predicted better at high learning rates. It is not clear if this behaviour would continue when predicting very long sequences, as the median length (i.e. duration) of mobility sequences available for testing was only 182 hours.
\\

\begin{figure}[h!]
\captionsetup{justification=centerlast}
\centering
\begin{subfigure}{.5\textwidth}
  \centering
  \includegraphics[width=0.8\linewidth]{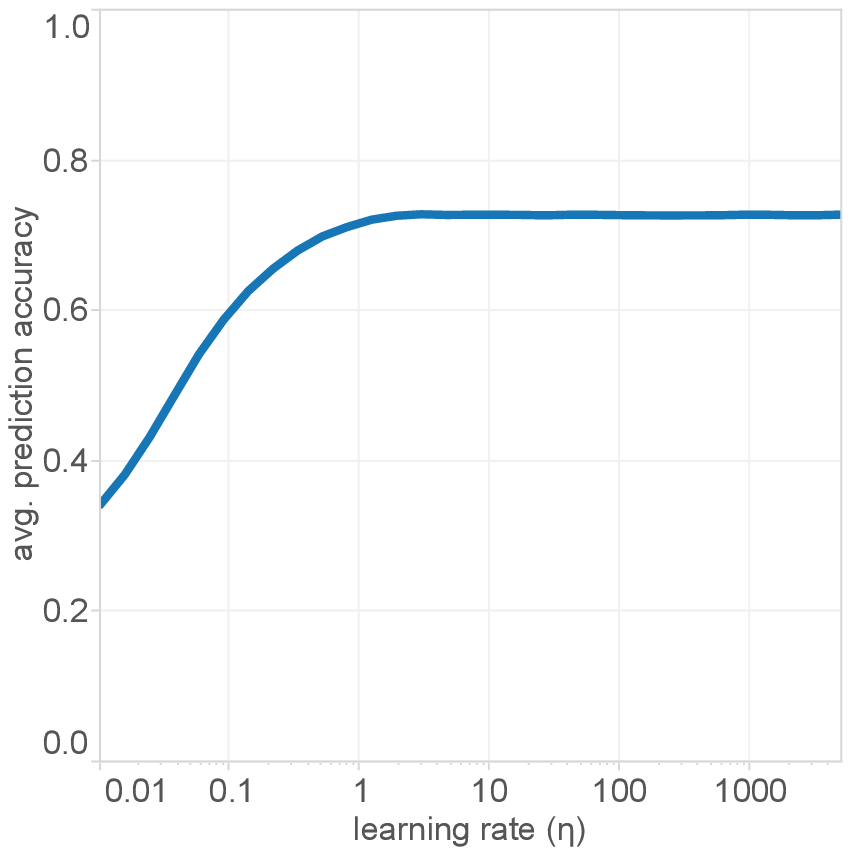}
  \caption{}
  \label{pred_per_eta}
\end{subfigure}%
\begin{subfigure}{.5\textwidth}
  \centering
  \includegraphics[width=0.8\linewidth]{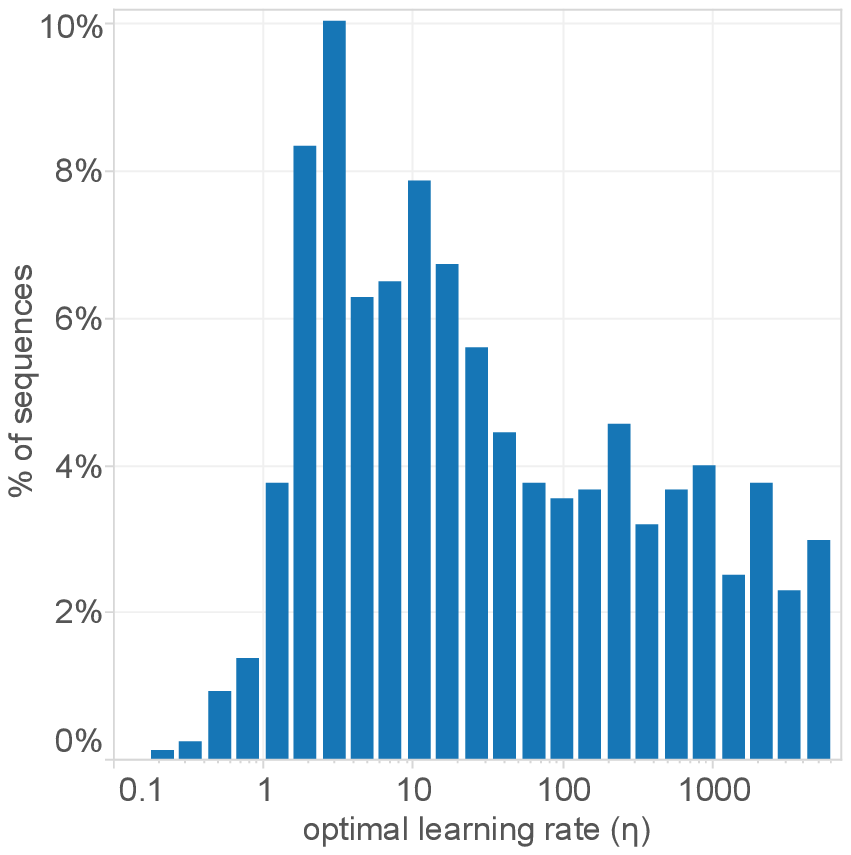}
  \caption{}
  \label{best_eta_dist}
\end{subfigure}
\caption{\textbf{Eta grid and optimal Eta.} (a) Average prediction accuracy as a function of $\eta$ for a logarithmic grid of 30 points equally log-spaced between $10^{-2}$ and $10^{3}$. The average prediction accuracy reaches a peak  at $\eta\approx 3$ and drops only very slightly after that, even for very high learning rates. This is due to the abundance of experts in the ensemble. Severely reducing an expert's weight after a even a single error in prediction with a large value of $\eta$ does not hurt the accuracy because there are many similar experts in the dataset. It is unclear however if this behaviour persists for longer sequences, since there were not enough long continuous sequences in the test set. (b) Distribution of the optimal learning rate $\eta$, i.e. the rate that achieves the best prediction accuracy. Values of $\eta$ are taken from a equally-spaced 30-point logarithmic grid. Most optimal $\eta$'s are quite large, indicating that the forecaster benefits from a strategy of immediate elimination of erroneous experts.}
\label{6ab}
\end{figure}

The non-adaptive versions with $\eta > 3$ achieved the best performance of $73\%$ average accuracy, but the adaptive versions achieved almost the same. Adaptation of $\eta$ will effectively cost a few steps to the forecaster, and since the value of $\eta$ beyond a certain threshold does not impact the result in our case, it does not seem to bring a benefit. However it is again unclear whether this result persists when predicting longer sequences. In addition to the adaptive/non-adaptive flavours of the forecaster, we tested changes to prediction accuracy when the Markov model of the user that is predicted is included in the expert ensemble. To respect causality, the user's own Markov model is of course changing as the successive locations of the user unfold in time, incorporating new data in its transition probabilities. The inclusion of the user's own data surprisingly seems to lower average predictability by almost two percentage points. This is true in both the adaptive and non-adaptive versions. 

\subsection*{S3. Transience of test sequences}
A symbolic sequence unfolding in time is stationary when the probability for the appearance of any given subsequence at a given position does not depend on the position, in the infinite time limit. The relative frequencies of transitions $P(X_m | X_{m-1}X_{m-2}\ldots X_{m-k})$, interpreted as the transition probabilities , must tend to  a limit value as the length of the sequence grows, for all $k$. In practice, the frequency of the patterns must be stable enough in the time scale of the problem under study. A human mobility sequence can appear quasi-stationary in this sense, when observed for a few weeks or months. But over many years most individuals will travel, change place of residence or in other ways exhibit significant shifts to their mobility patterns, during what could be called periods of transient mobility. Our roamer data captures examples of such transients, when users are visiting a foreign country, outside their regular mobility regime.

\begin{figure}[h!]
\captionsetup{justification=centerlast}
\centering
\includegraphics[width=1\linewidth]{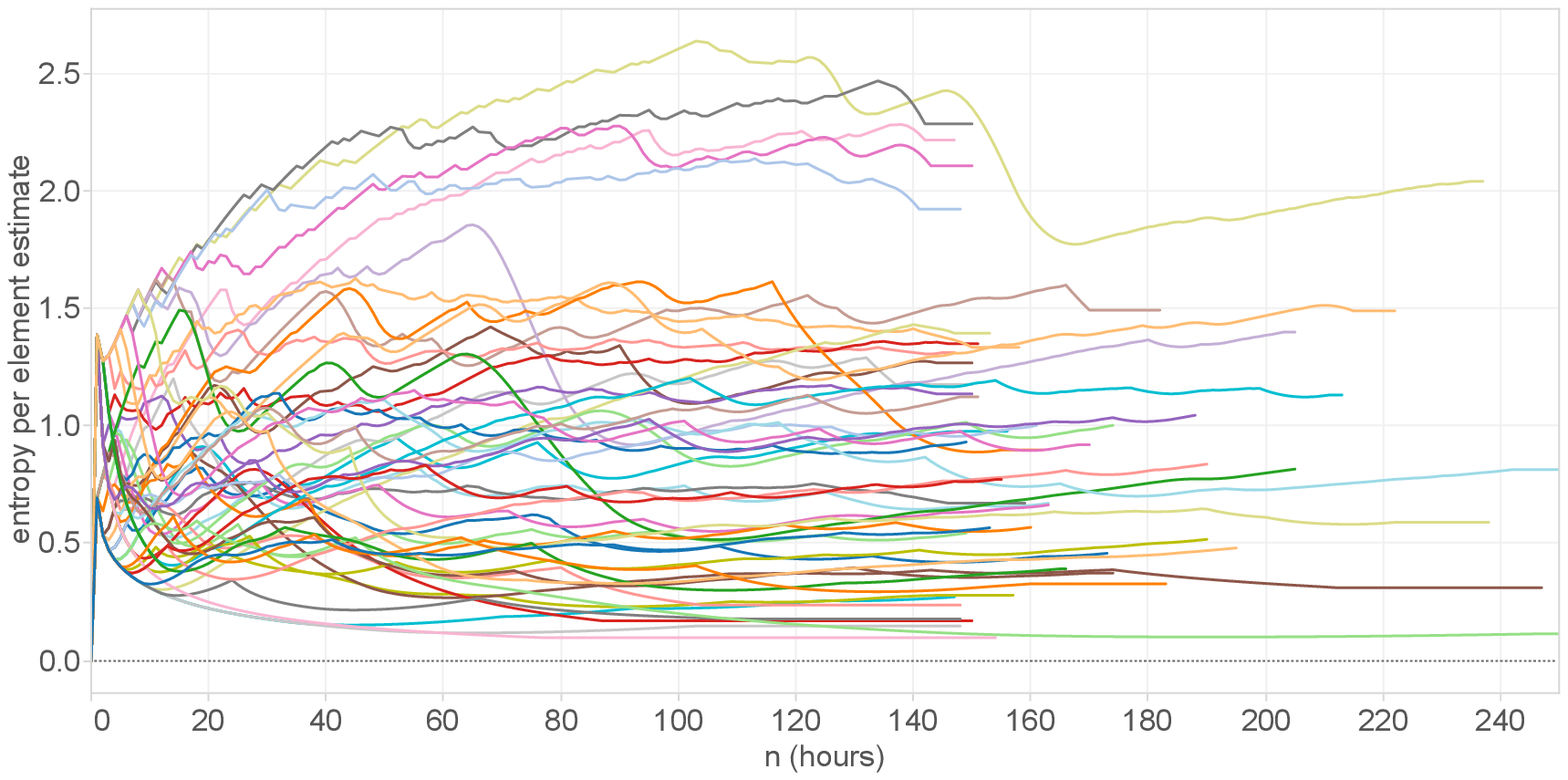}
\caption {\textbf{Entropy per element} estimate using the Lempel-Ziv estimator for a random sample of the test sequences. In an overwhelming majority of cases, the estimate has not stabilized before the sequence ends. This indicates that the mobility patterns that the sequence represents are not regular, or the sequence is too short to detect regularities.}
\label{convergence}
\end{figure}

To assess whether our test sequences can be considered quasi stationary we examine the standard Lempel-Ziv estimate of the entropy per element $H$ \cite{lempelziv, kontoyannis,barab_limits,ivory2013}:
\begin{equation}
\lim_{N} \frac{1}{N} \sum_{n=1}^{N} \frac{\Lambda_n}{\log{N}} = \frac{1}{H}
\end{equation}
Here $N$ is the length of the sequence, and $\Lambda_n$ is the length of the shortest subsequence starting at position $n$ that has not previously been encountered in the sequence. Since $H$ can be shown to converge to the true entropy per element of the sequence \cite{kontoyannis}, which for a stationary sequence is independent of the position, a fluctuating $H$ shows that the sequence cannot be considered quasi-stationary during the time interval of observation. Indeed, this is what we find for most sequences in our dataset, as can be seen in Supplementary Fig.~\ref{convergence} , where $H$ is plotted as a function of the position for a random sample of sequences from the test set. 

